# Controllable manipulation of superconductivity using magnetic vortices

J. E. Villegas[1] and Ivan K. Schuller[2]

[1]*Unité Mixte de Physique CNRS/Thales, 1 avenue A. Fresnel, 91767 Palaiseau, France and Université Paris Sud 11, 91405 Orsay, France*

[2]*Physics Department and Center for Advanced Nanoscience, University of California – San Diego, 9500 Gilman Drive, La Jolla, CA 92037, USA*

The magneto-transport of a superconducting/ferromagnetic hybrid structure consisting of a superconducting thin film in contact with an array of magnetic nanodots in the so-called "magnetic vortex-state" exhibits interesting properties. For certain magnetic states, the stray magnetic field from the vortex array is intense enough to drive the superconducting film into the normal state. In this fashion, the normal-to-superconducting phase transition can be controlled by the magnetic history. The strong coupling between superconducting and magnetic subsystems allows characteristically ferromagnetic properties, such as hysteresis and remanence, to be dramatically transferred into the transport properties of the superconductor.

Email: javier.villegas@thalesgroup.com

Email: ischuller@ucsd.edu

**Introduction.**

Recently, a large variety of artificial hybrid superconducting/ferromagnetic (S/F) systems have been investigated. Superconductivity and ferromagnetism are antagonistic long-range order phenomena. In hybrid systems, reduced dimensionality, confinement, and the intimate contact at the S/F interfaces enhance the competing interactions. As a result, a plethora of novel, unexpected behaviors have been found both in low- [1-4] and all-oxide high-$T_c$ S/F systems [5,6]. One interesting characteristics of certain S/F hybrids is that low-intensity external stimuli (e.g. magnetic fields) having moderate effects on the individual constituents may induce *dramatic* changes of the hybrid's properties, as they break a delicate balance resulting from the competing interactions. Ultimately, this provides opportunities for the fabrication of novel devices, such as superconducting memories [7].

An interesting possibility arises in hybrids in which the resistance of the S subsystem can be switched between two (or more) values, depending on the magnetic state of the F subsystem. An example of this is the so-called "superconducting spin-switch" effect, which has been observed in a variety of F/S/F heterostructures, both with high-[8,9] and low-$T_c$ [10-15] superconductors. The resistance switching is produced in some cases by spin accumulation effects [8,11,14,16], and in others by purely magnetostatic interactions due to the effect of the ferromagnet stray magnetic field on the superconductor [9,10,12,13,17]. The stray magnetic fields from ordered F nanostructures have also been used to produce a shift of the superconducting-to-normal (S-to-N) phase diagram of S/F hybrids, so that a superconducting state may be stabilized with an increased magnetic field [18].

In the present article, we investigate in detail a system in which the tunable, magnetic-history dependent stray magnetic field from a very dense F nanodot array controls the S-to-N transition of a superconducting thin film [19, 20]. One of the key properties of the system studied here is the possibility of changing the magnetic state of the F system (and thus the stray field profile) by applying external magnetic fields which are lower than (or comparable to) the superconducting critical fields. A strong coupling between the F and S system is obtained: small changes induced by the external applied field in the F-array magnetic state *simultaneously* show up as large changes of the resistance of the S. This way, details of the magnetic reversal of the F subsystem are transferred (and appear magnified) in the transport properties of the S one. Thus the reversible/irreversible magnetization induces a reversible/irreversible magneto-resistance of the superconductor, and consequently of the whole system. Moreover, for a given range of applied fields, the system can either be in the N or S state depending on the magnetic history, i.e. the superconductor becomes bistable. This produces an unusual, hysteretic, remanent magneto-resistance that reaches values of up to $10^5$ %, comparable or in excess of the "superconducting spin-switch" effects in F/S/F trilayers [8-15]. In order to induce these effects, we used very dense arrays of nanodots whose magnetic reversal takes place via the "vortex state" [21,22]. As we show below, the key property of nanodot magnetic vortices is that they virtually behave as magnetic dipoles whose polarity can be easily controlled, and therefore produce locally a very intense controllable magnetic field on the F film [23].

The paper is organized as follows. In section 2 we describe the sample fabrication and the magnetic properties of the F nanodot arrays. In section 3.1. we detail the effects of the stray magnetic field when an external magnetic field is applied perpendicular to the film plane. Section 3.2 addresses the central results of this paper, which appear when the magnetic field is applied in plane. Section 4 contains the conclusions.

## 1. Experimental.

Fe nanodots (20 nm thick) arrays were prepared on Si substrates using e-beam evaporation through nanoporous alumina masks. The Fe was capped with a 2 nm thick Au layer to prevent oxidation. After deposition of Fe/Au, the porous alumina mask was removed, to leave only de nanodot array on the substrate. The arrays have short-range hexagonal order (inset in Fig. 1 (b)) and a narrow distribution of nanodot diameters $\varnothing$ and interdot distances $d$ ($\Delta\varnothing \approx \Delta d \sim 10\%$). We studied different arrays with 55 nm<$\varnothing$<140 nm and 85 nm<$d$<180 nm. Further details of the fabrication and structural characterization can be found elsewhere [19,24]. The nanodot arrays were covered with a superconducting Al thin film (either 20 or 40 nm thick) using e-beam evaporation. A conventional 4-probe bridge was lithographed for magneto-transport measurements. The superconducting critical temperatures $T_c$ were in the range 1.28-1.42 K depending of Al thickness [20] (see an example of the SC transition in Fig. 1 (b)). Upper critical fields $H_{c2}$ obtained from magnetotransport imply a superconducting coherence length $\xi(0) \approx 45$ nm. The penetration depth $\lambda(0) \approx 250$ nm, estimated [25] from $T_c$ and the resistivity $\rho_{4.2K} \approx 6$ $\mu\Omega\cdot cm$ gives $\kappa(0) = \lambda(0)/\xi(0) \approx 5.5$, i.e. the Al films studied here are type-II superconductors. Magnetization measurements were made using a SQUID magnetometer, and magneto-transport experiments were carried out in a liquid He cryostat equipped with a superconducting magnet.

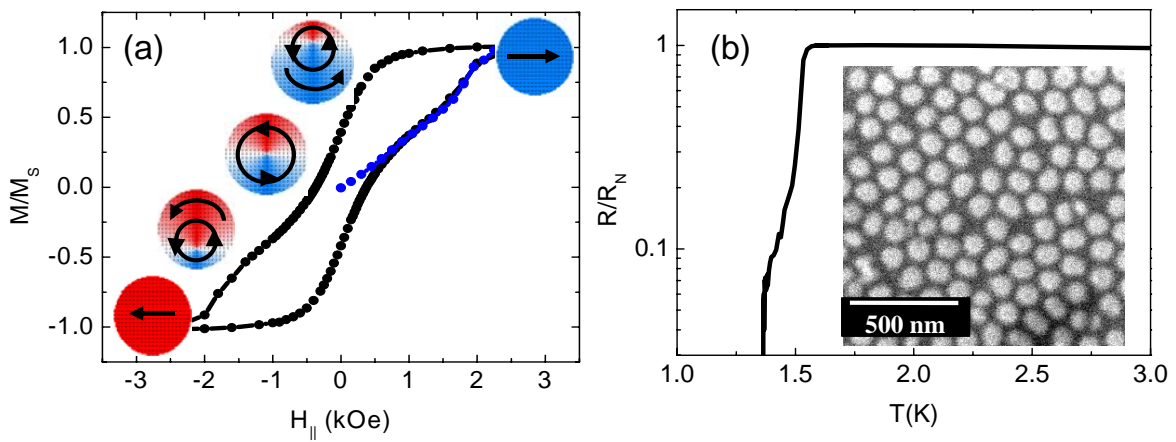

**Figure 1: (a)** Magnetization at T=6K of a sample with an array of magnetic nanodots with $\varnothing$=75±5 nm and $d$ =120±20 nm. The sketch is a cartoon of the magnetic reversal mechanism (see text). (b) R(T) of the as-grown sample, in zero applied external field. The inset shows a typical scanning electron microscopy image of the arrays.

Figure 1 (a) shows the typical hysteresis loop (with the external magnetic field $H$ applied in-plane) for an array in which magnetization reversal takes place via the magnetic "vortex state". This loop is representative of the behavior of arrays with nanodot diameters in the 75 nm<∅<140 nm range. The "pinching" around the coercive field is characteristic of the magnetization reversal via nucleation, displacement and annihilation of magnetic vortices [21-23]. In these dots, vortices nucleate around $H \approx -100$ Oe (nucleation field, $H_n$) and vortex annihilation occurs at $H \approx -1.4$ kOe (annihilation field, $H_a$) as $H$ is reduced from positive saturation [29]. Structural imperfections and variations from dot to dot induce an annihilation/nucleation field distribution, with full-width half-maximum ~0.5 kOe [29]. A cartoon of this reversal mechanism is shown as an inset in Fig 1 (a). In a magnetic vortex, the magnetization curls in-plane around a vortex core, where it points out-of-plane. Around the coercive field, the vortex core is centered within the nanodot. The "polarity" of the vortex (whether the magnetization in the core points "up" or "down") is set as vortex nucleation takes place. Once set, a relatively large out-of-plane $H$ is needed to flip the polarity [22]. The vortex core diameter is of the order of 15-20 nm in Fe nanodots, as expected from micromagnetic calculations [26] and confirmed by polarized Scanning Tunneling Microscopy [27] and neutron reflectometry [28]. A detailed study of the magnetic properties of these nanodot arrays can be found elsewhere [24,26,28,29].

## 2. Results and discussion.

### 2.1. Magneto-transport in out-of-plane ("perpendicular") external field.

The experiments in this section show that the superconducting/normal state the Al film is controlled by the distribution of vortex polarities in the array, which can be induced by applying relatively large perpendicular $H$ (as compared to the superconducting critical field).

Figure 2 (a) shows the normalized (to the normal state resistance $R_N$) magneto-resistance below $T_c$ of a sample with an array of nanodots with $\varnothing=75\pm5$ nm and $d=120\pm20$ nm, and the aluminum thickness $t_{Al}=20$ nm. The black curve (symmetric around $H=0$) was obtained for an array demagnetized prior to the $R(H)$ measurement. ($R(H)$ was measured first from $H=0$ to $H=150$ Oe and then, after a new demagnetizing cycle, from $H=0$ to $H=-150$ Oe). The red (blue) curve is measured after application and removal of a negative (positive) magnetizing $|H|=10\,\text{kOe}$. This induces a negative (positive) shift of $R(H)$, with the minimum resistance around $|H|\approx30$ Oe. The curve shape changes after magnetization, and an enhanced decrease in the resistance characterizing the N-to-S transition is observed, indicating a deeper superconducting state.

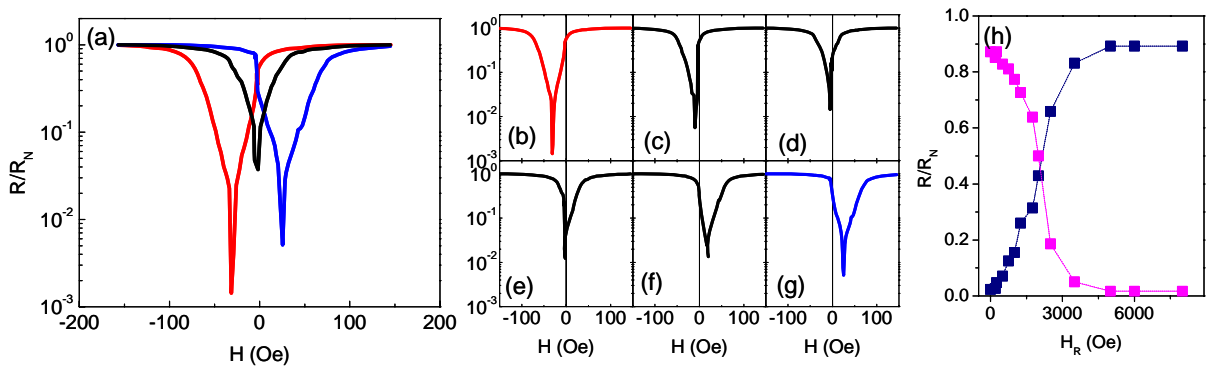

**Figure 2: (a)** Normalized resistance ($R_N$ the normal-state resistance) as a function of field applied perpendicular to the film plane, after a negative/positive magnetizing field (red/blue), and after demagnetization (black). At $T=0.9T_c$ and J=31.25 kA·cm$^{-2}$ **(b)-(g)** The same measurement after different applied field protocols (see text); $H_R=0$ (b), $H_R=0.5$ (c), $H_R=1.25$ (d), $H_R=2$ (e), $H_R=3.5$ (f), and $H_R=5$ kOe. **(h)** Normalized resistance ($R_N$ the normal-state resistance) for $H=-30$ Oe (blue) and $H=30$ Oe (magenta), as a function of the reversing field $H_R$.

Figures 2 (b)-(g) show $R(H)/R_N$ (with $R_N$ the normal-state resistance) after application of the following $H$ protocol: prior to the each $R(H)$ measurement, a negative magnetizing $H = -10$ kOe was followed by a reversed $H_R = 0$, 0.5, 1.25, 2, 3.5, and 5 kOe ( (b), (c), (d), (e), (f) and (g), respectively). $R(H)$ monotonously shifts from left to right as $H_R$ is increased. The evolution of $R(H)$ as a function of $H_R$ is analyzed in Figure 2 (h).

Figure 2 (h) shows the normalized resistance for $H = -30$ Oe ( $R(H = -30)/R_N$, blue) and for $H = 30$ Oe ( $R(H = 30)/R_N$, magenta), as a function of the $H_R$ (note that $R(H)$ exhibits a minimum at $H \sim \pm 30$ Oe after application and removal of a high positive/negative magnetizing field, see figure 2 (a)). The resistances shown in Fig. 2 (h) are obtained from a series of curves as in Fig. 2 (b)-(g). The trends in Fig. 2 (h) are connected to the shift and shape of the $R(H)$ curves as a function of $H_R$. The crossing of the two curves in Fig. 2 (h) at $H_R \approx 1.75$ kOe inplies a symmetric $R(H)$ around $H = 0$. The saturation for $H_R \geq 5$ kOe implies that beyond this field $R(H)$ no longer shifts with increasing $H_R$.

The behavior described above is connected to the magnetic-history-dependent stray magnetic field from the vortices in the nanodot array. The application and removal of a sufficiently strong out-of-plane $H$ produces a complete polarization of the array; i.e. in the remanent state all vortices in the array have the same polarity [21,23]. After demagnetization, a balanced distribution of vortex polarities in the array is obtained (50 % "up", 50 % "down") [21, 23]. From this, we can calculate the magnetic field induced by the array of magnetic nanodots on the S film i) after application and removal of a magnetizing field, and ii) after demagnetization.

Fig. 3 (a) shows the out-of-plane magnetic field component created at the Al film plane by a single vortex core with "up" polarity (positive magnetization), as a function of the distance $r$ from its center, $H_\perp(r)$. Figure 3 (b) and (c) display the out-of-plane component of the magnetic field at the Al film plane $H_\perp(x,y)$ (with $x,y$ the in-plane coordinates) from magnetized (all the magnetic vortices point "up") and demagnetized arrays, respectively. For the latter, we assumed a random distribution of polarities. $H_\perp$ was calculated using basic magnetostatics [30]. A $9 \times 9$ hexagonal nanodot array with the same interdot distances as our samples was considered for (b) and (c). We used [19] the expression $M_\perp[r] = M_S(\Theta[1-r] + \Theta[r-1](s+1-r)/s)$ for the out-of-plane magnetization within the vortex cores, where $r$ is the distance from the center of the vortex ($r \leq 10$ nm), $\Theta[x]$ is the Heavyside's step function, $M_S = 1.65$ kOe is the saturation magnetization of Fe and $s = 8.97 \pm 0.13$ was obtained by fitting the expression to the experimental profile [27] of the magnetization within the core of Fe magnetic vortices.

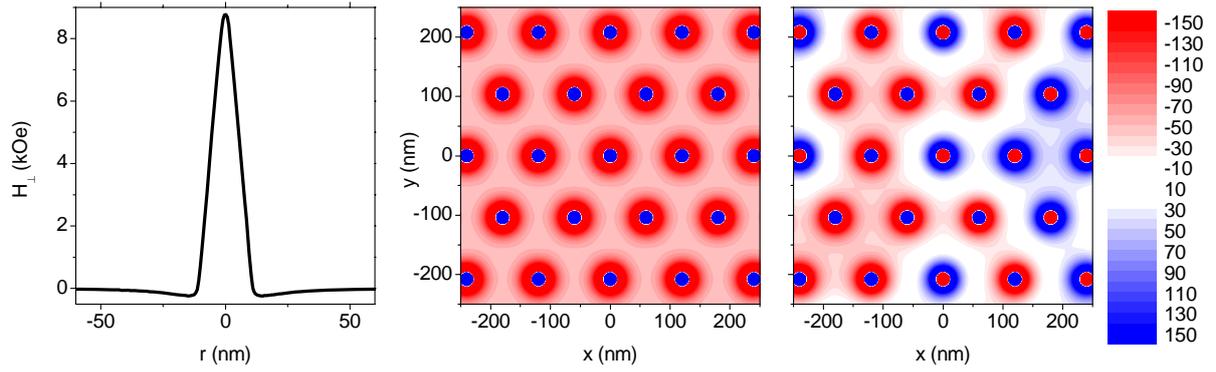

**Figure 3:** (a) Out-of-plane component of the magnetic field induced by a magnetic vortex as a function of the distance from the center of the core $r$. (b) Contour plot of the out-of-plane component of the magnetic field at the film plane from a completely "up" polarized array ($d$=120 nm) of magnetic vortices (c) the same for a demagnetized array (random distribution of "up" and "down" polarities). The color grade scale saturates at $|H_\perp|$=150 Oe.

As shown in Figure 3 (a), just above the vortex core the stray field is very intense (~ 9 kOe) as compared to the upper perpendicular critical field ($H_{c2}$ ~ 100 Oe at this temperature) [31]. Away from the vortex core, the stray field becomes negative (points in the opposite direction to the magnetization), and its magnitude decreases from one hundred to a few tens of Oe. The calculations for the polarized array (figure 3 (b)) show that the field is minimum in between the dots, where $H_\perp \approx 45$ Oe. For the demagnetized array, however, the field in regions between vortices with opposite polarities can be as low as ~2 Oe (figure 3 (c)). These calculations provide the understanding of the observations in Fig. 2 (a). After saturation (blue and red curves in fig. 2 (a)) the sample is close to the normal state in the absence of a external magnetic field $H$, because the stray field from the nanodots is intense enough to drive the sample into the normal state. When $H$ is applied parallel to the magnetization of the vortex cores, the compensation of the stray magnetic field in between the nanodots induces the transition into the superconducting state, similarly to what is observed for larger, fully magnetized dots [18]. Note the discrepancy between the experimental $H$ at which compensation occurs (~ 30 Oe) and the calculated minimum stray field in between the nanodots (~ 45 Oe). This could be due to the incomplete polarization of the array, since dipolar interactions between nanodots are not negligible for $d<2\varnothing$ [26], and therefore it is possible that in the remanent state the array is not fully polarized. Moreover, the calculations in Figure 3 (b)-(c) are for the field induced by the nanodot array, that does not take into account Meissner screening currents in the S film which may reduce the effective field in between the nanodots, leading to a lower external cancellation field. For a more accurate quantitative description of the cancellation between stray and external fields, a detailed calculation within the frame of Ginzburg-Landau theory [32] would be required. Note that our situation is different than in most of the calculations in the literature, because the stray magnetic field here is very intense and inhomogeneous over a length scale of a few tens of nm, shorter than the superconducting characteristic lengths $\xi(0) \approx 45$ nm and $\lambda(0) \approx 250$ nm.

When the array is demagnetized, superconductivity nucleates in the regions in between vortices with opposite polarity, where the stray magnetic field is very low. Percolation of superconducting currents along these regions explains the low resistance at $H=0$ after demagnetization (black curve in figure 2 (a)). Note, however, that the resistance is higher than when $H$ compensates the stray field (for $|H|$~30 Oe after magnetization, red/blue curves). In the latter case, superconductivity nucleates in between the nanodots *all across the*

*array*. However, in the demagnetized state, only a fraction of the space in between nanodots (20 to 36 %, depending on the degree of order of the array [19]) is subject to a stray magnetic field below 30 Oe. This explains the higher resistance observed in the demagnetized state around $H=0$ as compared to the resistance at the compensation field ($|H|\sim 30$ Oe) after magnetization.

The trends in Fig. 2 (h) can be understood considering that, after application and removal of a negative saturating $H$ which induces the same polarity for all the vortices in the array, the application of a positive $H_R$ switches the polarity of some. Due to the distribution nanodot diameters, a distribution of polarity switching fields is expected [29]. The crossing of the two curves in Figure 2 (h) at $H_R \sim 1.75$ kOe implies that this field reverses the polarity of around 50% of the vortices in the array, leading to a "balanced" distribution as the one in fig. 3 (c). The saturation of the curves above 5 kOe implies that this field suffices to flip the polarity of 100% of the vortices.

*2.2. Magneto-transport with an in-plane external field applied*

Figure 4 (a) shows in-plane $M(H)$ (for the same sample discussed in the previous section) after demagnetization ("virgin" curve), as $H$ is cycled from $H=0$ up to different maximum $H_{max}$ and back to $H=0$. For $H_{max}$ smaller than $\sim 1.5$ kOe (blue curve) the magnetization is reversible, i.e if $H$ is swept up and down between zero and this threshold value, no change in the $M(H)$ nor remanence are observed. Irreversibility and remanence appear above $H_{max} \sim 1.5$ kOe (green curve). The remanence increases with increasing $H_{max}$ until saturation around $H_{max} = 2.5$ kOe. This behavior is characteristic of arrays of magnetic vortices [29]. The threshold $H \sim 1.5$ kOe lies in the range of vortex annihilation fields for these arrays $H_a \sim 1.4$ kOe [29]. The magnetization is reversible below this threshold $H$ because the magnetization changes are produced by the reversible displacement of the vortex core [22,29]. Above this

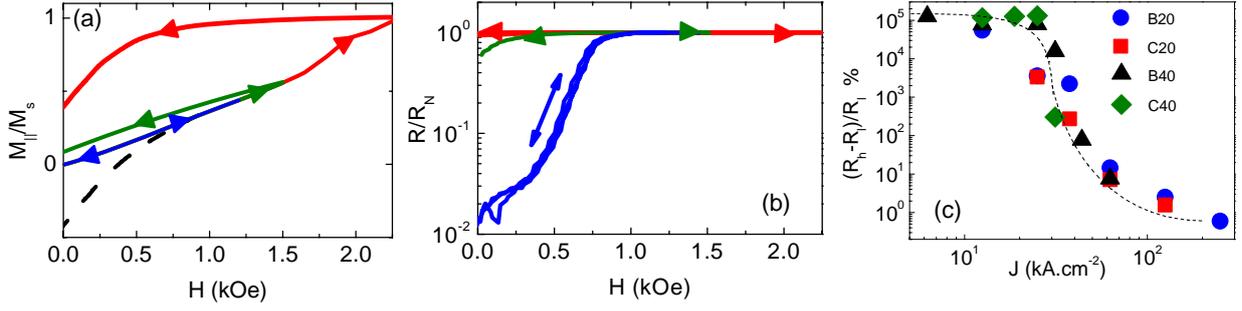

**Figure 4 (a)**: Magnetization at T=6K of a sample with an array of magnetic nanodots with ⌀=75±5 nm and $d$ =120±20 nm. Coloured curves are for the field excursions described in the text $H=0$ →$H_{max}$ →$H=0$ after demagnetization, with $H_{max}$=1.3 kOe (blue), $H_{max}$=1.6 kOe (green) and $H_{max}$=2.5 kOe (red). **(b)** Normalized resistance ($R_N$ is the normal-state one) as a function of the field applied parallel to the film plane, $T=0.89T_c$ and J=25 kA·cm$^{-2}$, for the same field cycles as in (a). **(c)** Remanent magneto-resistance (see text for definition) vs. injected current density $J$. Different colors or for different samples B20 (⌀=75±5 nm and $d$ =120±20 nm and aluminium thickness $t_{Al}$=20 nm), C20 (⌀=140±20 nm, $d$ =180±40 nm and $t_{Al}$=20 nm), B40 ⌀=75±5 nm and $d$ =120±20 nm and aluminium thickness $t_{Al}$=40 nm) and C40 (⌀=140±20 nm, $d$ =180±40 nm and $t_{Al}$=40 nm) threshold, the annihilation (with increasing $H$) and nucleation (decreasing $H$) of vortices give rise to the irreversibility [29].

Figure 4 (b) shows the $R(H)$ at $T=0.89T_C$, with $H$ in-plane. Measurements were done with the same $H$ cycling sequence as in Fig. 4 (a). The same reversible and irreversible regimes as for $M(H)$ are observed. After demagnetization, the resistance around $H=0$ is two orders of magnitude below the normal state one. Application of $H$ gradually drives the sample into de normal state. For $H_{max}$ lower than 1.5 kOe (blue curve), the low resistance state is recovered upon removal of $H$. Within this range of fields, $R(H)$ is reversible and remains constant independently of the number of field cycles. However, for $H_{max}$ above 1.5 kOe a remanent resistance appears upon removal of the applied field, which increases with increasing $H_{max}$ until saturation around $H_{max}=2.5$ kOe. In summary i) for a range of $H$ around $H=0$ the sample is either superconducting or normal depending on the magnetic history, ii) the switching between these states occurs as $H$ is cycled around the vortex annihilation fields, iii) the remanence and the reversible/irreversible magnetization characters are transferred into the magneto-transport of the superconductor.

Figure 4 (c) shows the current density ($J$) dependence of the remanent magnetoresistance defined as $MR \equiv (R_N - R_h / R_l)$, where $R_N$ and $R_l$ are respectively the normal-state and the lowest resistance for $H = 0$. Data for different Al thickness and nanodot sizes are shown (see figure caption). All samples behave similarly: the remanent magnetoresistance saturates at around $10^5$ % at low current levels, and gradually decreases to 1% for currents above 100 kA.cm$^{-2}$. Note that the magneto-resistance at low currents is comparable or in excess of that observed in trilayer S/F/S systems from the so-called "spin switch effects" [8-17], for which the MR is defined in a similar way.

Further correlations between magnetization and resistance are shown in Figure 5. (a) to (i) depict a series of $R(H)$ measured as $H$ is cycled from positive saturation $H_S = 3$ kOe to a reversal $H_R \leq 0$, and then swept back to positive saturation ($H_S \rightarrow H_R \rightarrow H_S$; minor loops). Similar $M_\parallel(H)$ cycles were independently measured (not shown). The magneto-resistance is strongly hysteretic. Cycles with $|H_R|$ in the 1-1.25 kOe range, (e) and (f), lead to the lowest resistance (superconducting state), while those with larger or smaller $|H_R|$ lead to a high-resistance state. In particular, the sample is in the normal-state around $H = 0$ for $H_R > -0.25$ kOe (figures 5 (a)-(b)) and for $H_R < 2$ kOe (figure 5 (i)). The behavior observed in figures 4 and 5 appear only for samples in which the nanodots' magnetic reversal takes place via the vortex state. Experiments for similar samples with single magnetic domain nanodots do not show bistability and magnetic hysteresis as described above [19].

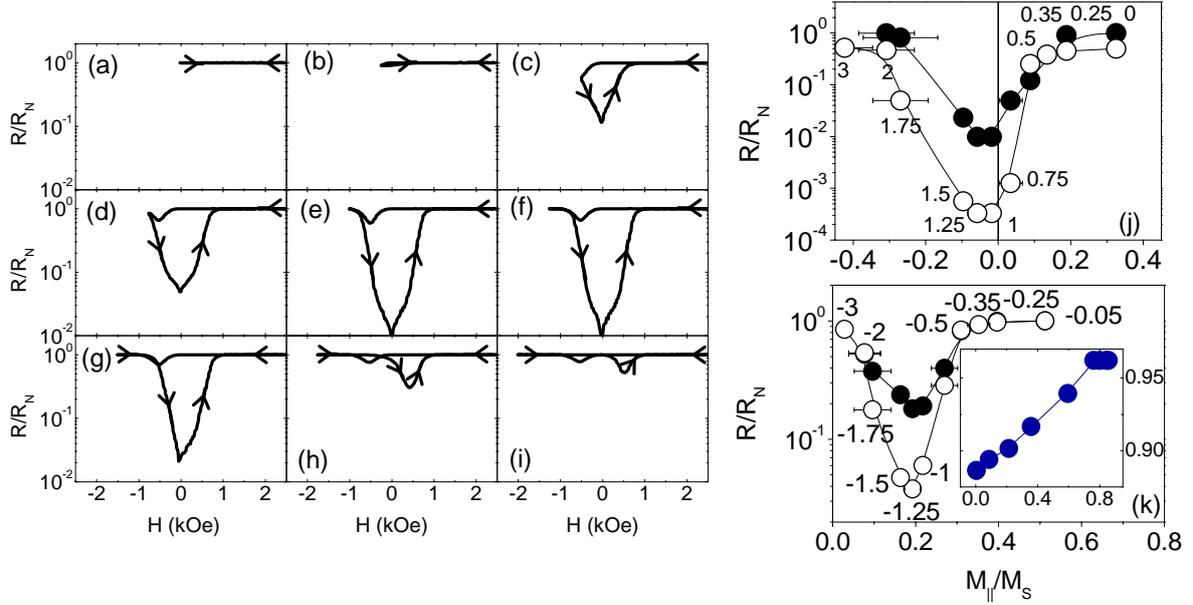

**Figure 5: (a)-(i)** Normalized resistance ($R_N$ is the normal-state one) as a function of the field applied parallel to the film plane, $T=0.89 T_c$ and $J=25$ kA·cm$^{-2}$, for the field cycles $H_s \rightarrow H_R \rightarrow H_s$ described in the text, with $H_R$=0, -0.35, -0.5, -0.75, -1, ,-1.25, -1.5, -1.75 and -2 for (a), (b), (c), (d), (e),(g),(h) and (i) respectively. **(j)** Normalized resistance vs. normalized *in-plane* magnetization ($M_S$ saturation magnetization) at *zero applied field*. Solid (hollow) circles for samples with $\varnothing$=75±5 nm and $d$=120±20 nm and aluminium thickness $t_{Al}$=20 nm ($t_{Al}$=40 nm ). Each point corresponds to a $|H_R|$ indicated by the numbers close to each point. Lines are guides to the eye. **(k)** Same as in (j) for with an applied field $H$=0.5 kOe. The inset shows the same experiment for a sample with $\varnothing$=55±5 nm, $d$=85±20 nm and $t_{Al}$=20 nm, in which nanodots are magnetic single domains.

Combining $R(H)$ and $M_{||}(H)$ we obtained $R(M_{||})$ for any field $H$. Figure 5 (j) and (k) show respectively $R(M_{||})$ at *zero applied magnetic field* $H=0$ and at $H=0.5$ kOe (the field at which a small minima is observed after negative saturation) for two different Al thicknesses on the same array: 20 nm (black circles) and 40 nm (white circles). The labels indicate $|H_R|$. This figure describes the effect of the dots' magnetization on the hysteretic magneto-resistance. An *in-plane* magnetization $|M_{||}/M_S| > {\sim}0.4$ drives the sample into the normal-state independently of the magnetic history or external applied field. However, $M_{||}$ cannot account by itself for the switching between the different states in Fig. 4 (a)-(i). This is evidenced by the strong asymmetry in figure 5 (j), where $R(M_{||}) \neq R(-M_{||})$ around

$|M_{||}/M_S| \approx 0.05 - 0.15$. Within this range, $R(-M_{||})$ is two to four orders of magnitude lower than $R(M_{||})$. Further evidence arises from figure 5(k), where a non monotonic dependence of $R$ on $M_{||}$ is observed. This is in contrast to the behavior observed in samples with arrays of slightly smaller (∅<60 nm) single domain in-plane nanodots [19]. For these, the resistance increases monotonically with increasing magnetization, as shown in the inset of figure 5 (k). All this suggests that, besides the *in-plane* magnetization $M_{||}$, the *out-of-plane* vortex core magnetization $M_\perp$ must play a significant role. Note in addition (see labels in figs. 5 (j)-(g)) that the lowest resistance state is achieved with cycles in which $|H_R| \sim 1 - 1.5$ kOe . These fields are within the distribution of vortex annihilation fields found for similar arrays (∅=67 nm), for which the annihilation fields $|H_a| \sim 1.4$ kOe with a full-width at half-maximum $\Delta H_a \sim 0.5$ kOe [29]. For cycles with $|H_R|$ out of this range, the sample remains close to or in the normal state. This implies that the high- to low-resistance state transition is triggered as some fraction of the magnetic vortices are annihilated around $|H_R|$ and re-nucleate when $H$ is swept back to $H = 0$. Contrary to this, the high resistance (normal) state is achieved with cycles in which i) $|H_R|$ is below the annihilation field or ii) $|H_R|$ is high enough to annihilate *all* vortices.

Experiments in section 3.1 showed that the stray magnetic field produced by the nanodots' out-of-plane magnetization $M_\perp$ is intense enough to suppress superconductivity on top of the vortex cores, but allows nucleation of superconductivity in between them according to the vortex polarity distribution. From this, the connection between the hysteretic magneto-resistance and $M_\perp$ can be understood if we consider that the different in-plane $H$ cycles set different vortex polarity distributions in the array. In principle, if the system has perfect reflection symmetry with respect the film plane, a balanced random distribution of polarities

(Figure 3 (c)) is expected when the in-plane $H$ is withdrawn after saturation, and vortices nucleate. However, minor unavoidable misalignments between $H$ and the array plane will induce a uniform polarity in the array (similar as in Figure 3 (b)). This was found with micromagnetic calculations for these arrays (using the OOMMF code [33] and same parameters as in [26,29]). Simulations for arrays of 22 dots showed that 85 % of them nucleate with the same polarity as $H$ is reduced from saturation if the misalignment is as little as 4 degrees off the array plane. From this, we can understand the behaviors found if Figure 4 and 5. Field protocols for which the in-plane $H$ exceeds at large the range of vortex annihilation fields produce the annihilation of *all* the vortices in the array. Most of these nucleate with equal polarity upon withdrawal of $H$ (situation similar to Figure 3 (b)) yielding a high-resistance state around $H = 0$. Protocols in which $H$ is cycled within the range of vortex annihilation fields will produce to the annihilation of *only a fraction* of the vortices. Upon reduction of $H$, most of these will re-nucleate with the polarity imposed by the external applied field, while those that where not annihilated will keep their prior polarity. Eventually, this process will result in a situation as the one in Figure 3 (c) (depending on the array state prior to the $H$ cycle and the polarity induced by the external applied field), producing the lowest resistance state. Since the polarity distributions only change through annihilation/nucleation processes, the low-resistance state is robust against $H$ cycles, and reversible behavior is observed (figure 4 (b)) as long as the applied fields are below the annihilation fields.

4. **Conclusions**

We have investigated a S/F hybrid system in which the stray magnetic fields from a very dense magnetic array controls the superconducting-to-normal transition of a superconducting thin film. Different magnetic states can be induced in the array upon application and withdrawal of relatively high out-of-plane magnetic fields, which produces a shift of the superconducting-to-normal phase boundary. When magnetic fields are applied in-plane, the magnetic state of the array can be controlled by application of lower fields, which produces a strong coupling between the transport and magnetic response of the sample. This results in an unusual, hysteretic, remanent magnetoresistance, in which the different reversible/irreversible regimes of the magnetic arrays are imprinted into the superconductor. The key ingredient to this behavior is the controllable out-of-plane magnetization of the magnetic vortices within the array nanodots.

5. Acknowledgments

We thank C.-P. Li for sample fabrication and OOMF simulations. Work at UCSD supported by the NSF. French RTRA "Supraspin" and ANR "Superhybrids-II" grants are acknowledged.

6. **References.**